\documentstyle[12pt,psfig]{article}
\def\pew{P_{EW}}
\def\pewc{P_{EW}^C}
\newcommand{\thspace}{\kern.08333em}

\def \beq{\begin{equation}}
\def \eeq{\end{equation}}
\def \beqn{\begin{eqnarray}}
\def \eeqn{\end{eqnarray}}

\def \s{\sqrt{2}}

\def \v#1#2{V_{#1#2}}

\def\rly#1{\mathrel{\raise.3ex\hbox{$#1$\kern-.75em\lower1ex\hbox{$\sim$}}}}

\def\gsim{\rly>}

\def \olambdai{ {\cal O} (\lambda)}
\def \olambda#1{ {\cal O}(\lambda^#1)}
\begin{document}
\rightline{TECHNION-PH-95-11}
\rightline{UdeM-GPP-TH-95-25}
\rightline{EFI-95-15}
\rightline{hep-ph/9504327}
\rightline{April 1995}
\bigskip
\bigskip
\centerline{\bf ELECTROWEAK PENGUINS AND TWO-BODY $B$ DECAYS}
\bigskip
\centerline{\it Michael Gronau}
\centerline{\it Department of Physics}
\centerline{\it Technion -- Israel Institute of Technology, Haifa 32000,
Israel}
\medskip
\centerline{and}
\medskip
\centerline{\it Oscar F. Hern\'andez\footnote{e-mail:
oscarh@lps.umontreal.ca} {\rm and} David London\footnote{e-mail:
london@lps.umontreal.ca}}
\centerline{\it Laboratoire de Physique Nucl\'eaire}
\centerline{\it Universit\'e de Montr\'eal, Montr\'eal, PQ, Canada H3C 3J7}
\medskip
\centerline{and}
\medskip
\centerline{\it Jonathan L. Rosner}
\centerline{\it Enrico Fermi Institute and Department of Physics}
\centerline{\it University of Chicago, Chicago, IL 60637}
\bigskip
\centerline{\bf ABSTRACT}
\medskip
\begin{quote}

We discuss the role of electroweak penguins in $B$ decays to two light
pseudoscalar mesons. We confirm that the extraction of the weak phase
$\alpha$ through the isospin analysis involving $B\to\pi\pi$ decays is
largely unaffected by such operators. However, the methods proposed to
obtain weak and strong phases by relating $B\to\pi\pi$, $B\to\pi K$ and
$B\to KK$ decays through flavor SU(3) will be invalidated if electroweak
penguins are large. We show that, although the introduction of electroweak
penguin contributions introduces no new amplitudes of flavor SU(3), there
are a number of ways to experimentally measure the size of such effects.
Finally, using SU(3) amplitude relations we present a new way of measuring
the weak angle $\gamma$ which holds even in the presence of electroweak
penguins.

\end{quote}
\newpage

\centerline{\bf I. INTRODUCTION}
\bigskip

The $B$ system is the ideal place to measure the phases of the
Cabibbo-Kobayashi-Maskawa (CKM) matrix. The weak phases $\alpha$, $\beta$
and $\gamma$ can be measured in numerous ways through asymmetries and rate
measurements of various $B$ decays \cite{CPreview}. Ultimately it will be
possible to verify the relation $\alpha=\pi-\beta-\gamma$, predicted within
the Standard Model.

The conventional method for obtaining the angle $\alpha$ is through the
measurement of the time-dependent rate asymmetry between the process $B^0
\to \pi^+\pi^-$ and its CP-conjugate. This assumes that the decay is
dominated by one weak amplitude -- the tree diagram. However, there is also
a penguin contribution to the decay, which has a different weak phase than
the tree diagram. This introduces a theoretical uncertainty into the
extraction of $\alpha$. Fortunately, this uncertainty can be removed by the
use of isospin \cite{GL}. The two final-state pions can be in a state with
$I=2$ or $I=0$. But the penguin diagram, which is mediated by gluon
exchange, contributes only to the $I=0$ final state. Thus, by isolating the
$I=2$ component, one can isolate the tree contribution, thereby removing
the uncertainty due to the penguin diagrams. This can be done through the
use of an isospin triangle relation among the amplitudes for
$B^+\to\pi^+\pi^0$, $B^0\to\pi^+\pi^-$ and $B^0\to\pi^0\pi^0$. By measuring
the rates for these processes, as well as their CP-conjugate counterparts,
it is possible to isolate the $I=2$ component and obtain $\alpha$ with no
theoretical uncertainty. The crucial factor in this method is that the
$I=2$ amplitude is pure tree and hence has a well-defined CKM phase.

Recently, it was proposed that the phases of the CKM matrix could be
determined through the measurement of various decay rates of $B$ mesons to
pairs of light pseudoscalars \cite{BPP,PRL,PLB}. This was based on two
assumptions: (i) a flavor SU(3) symmetry \cite{DZ,SW,Chau} relating
$B\to\pi\pi$, $B\to\pi K$ and $B\to KK$ decays, and (ii) the neglect of
exchange- and annihilation-type diagrams, which are expected to be small
for dynamical reasons. For example, it was suggested that the weak phase
$\gamma$ (equal to Arg$(V_{ub}^*)$ in the Wolfenstein parametrization
\cite{LW}), could be found by measuring rates for the decays $B^+ \to \pi^0
K^+$, $B^+ \to \pi^+ K^0$, $B^+ \to \pi^+ \pi^0$, and their
charge-conjugate processes \cite{PRL}. The $\pi K$ final states have both
$I=1/2$ and $I=3/2$ components. The key observation is that the
gluon-mediated penguin diagram contributes only to the $I=1/2$ final state.
Thus, a linear combination of the $B^+ \to \pi^0 K^+$ and $B^+ \to \pi^+
K^0$ amplitudes, corresponding to $I = 3/2$ in the $\pi K$ system, could be
related via flavor SU(3) to the purely $I = 2$ amplitude in $B^+ \to \pi^+
\pi^0$, permitting the construction of an amplitude triangle. The
difference in the phase of the $B^+ \to \pi^+ \pi^0$ side and that of the
corresponding triangle for $B^-$ decays was found to be $2 \gamma$. Taking
SU(3) breaking into account, the analysis is unchanged, except that one
must include a factor $f_K/f_\pi$ in relating $B\to\pi\pi$ decays to the
$B\to\pi K$ decays \cite{GHLR}. The weak phase $\gamma$ can also be
extracted in an {\it independent} way, along with the CKM phase $\alpha$
and all the strong final-state phases, by measuring the rates for another
set of 7 decays, along with the rates for the charge-conjugate decays
\cite{PLB}. (SU(3)-breaking effects are discussed in \cite{GHLR}.)
This method also relies on the SU(3) relation between the $I=3/2$ $\pi
K$ amplitude and the $I=2$ $\pi\pi$ amplitude.

The crucial ingredient in the above analyses is that the penguin is
mediated by gluon exchange. However, there are also electroweak
contributions to the processes $b\to sq{\bar q}$ and $b\to dq{\bar
q}$, consisting of $\gamma$ and $Z$ penguins and box diagrams. (From
here on, we generically refer to all of these as ``electroweak
penguins.'') Since none of the electroweak gauge bosons is an
isosinglet, these diagrams can affect the above arguments. For the
$B\to\pi\pi$ isospin analysis, the result is that the $I=2$ state will
no longer have a well-defined weak CKM phase. For the $B\to\pi\pi/\pi
K$ analyses, in the presence of electroweak penguins there are no
longer triangle relations among the $B\to \pi K$ and $B\to\pi\pi$
amplitudes. Theoretical estimates \cite{DH} have indicated that
electroweak penguins are expected to be relatively unimportant for
$\pi\pi$. However, they are expected to play a significant role in the
$\pi K$ case, introducing considerable uncertainties in the extraction
of $\gamma$ as described above.

The purpose of the present paper is to examine the role of electroweak
penguins in all $B\to PP$ decays, where $P$ denotes a light pseudoscalar
meson. We wish to address the following questions:

(1) To what $B$ decays do electroweak penguins contribute?

(2) Can one obtain information on their magnitude directly from the data?

(3) Can one extract weak CKM phases in the presence of electroweak
penguins?

We answer the first question by including the electroweak penguin
contributions in a general graphical description of all $B\to PP$
amplitudes, which was shown to be a useful representation of flavor SU(3)
amplitudes \cite{BPP}.

The second question is answered in the affirmative. An explicit calculation
of electroweak penguins \cite{DHT} suggests that they could dominate in
decays of the form $B_s \to (\phi~{\rm or}~\eta) + (\pi~{\rm or}~\rho)$. We
find that there are additional measurements which are indirectly sensitive
to such contributions.

As to the third question, we find that it is indeed possible to obtain
information about the CKM angle $\gamma$, even in the presence of
electroweak penguins. While the method proposed makes use of a considerably
larger number of measurements than the original simple set proposed in
\cite{BPP,PRL,PLB}, there is no difficulty {\it in principle} in obtaining
the necessary information from experiment alone. Whether these measurements
are feasible {\it in practice} in the near term is another story, which we
shall address as well. The four amplitudes for different charge states in
$B \to \pi K$ decays satisfy a quadrangle relation dictated entirely by
isospin. When sides are chosen in an appropriate order, we find that one
diagonal of the quadrangle is related to the rate for $B_s \to \pi^0 \eta$,
so that (up to discrete ambiguities) the quadrangle is of well-defined
shape. The difference between the other diagonal and the corresponding
quantity for charge-conjugate processes, when combined with the rate for
$B^+ \to \pi^+ \pi^0$, provides information on $\sin \gamma$.

We discuss general aspects of electroweak penguins in Sec.\ II, with
particular emphasis on estimates of the size of such effects. In Sec.\ III
we examine the electroweak penguin contributions to $B\to PP$ decays. The
quadrangle for $B \to \pi K$ decays is treated in Sec.\ IV. Experimental
prospects are noted in Sec.\ V, while Sec.\ VI summarizes.

\newpage 
\centerline{\bf II. ELECTROWEAK PENGUINS: GENERAL CONSIDERATIONS}
\bigskip

\leftline{\bf A. How big are electroweak penguins?}
\bigskip

The standard penguin diagram involves a charge-preserving, flavor-changing
transition of a heavy quark to a lighter one by means of a loop diagram
involving a virtual $W$ and quarks, and emission of one or more gluons. The
penguin diagrams involving $\bar b \to \bar d$ transitions change isospin
by 1/2 unit, while $\bar b \to \bar s$ transitions leave isospin invariant.

Penguin diagrams in which the $\bar b \to \bar q$ system is coupled to
other quarks through the photon or $Z$ (or through box diagrams involving
$W$'s) instead of through gluons have more complicated isospin properties.
There will be contributions in which the additional quark pair is isoscalar
(as in the conventional penguin graphs), but others in which it is
isovector.

The importance of electroweak penguin (EWP) diagrams was realized in the
calculation of the parameter $\epsilon'/\epsilon$ describing direct CP
violation in $K_L \to \pi \pi$ \cite{KEWP}. That parameter requires an
imaginary part of the ratio $A_2/A_0$, where the subscript denotes the
isospin $I_{\pi \pi}$ of the $\pi \pi$ system. The EWP can provide an
$I_{\pi \pi} = 2$ contribution, whereas the conventional penguin cannot.
The numerical importance of the EWP diagram involving $Z$ exchange is
enhanced by a factor of $m_t^2/M_Z^2$ \cite{GSW}.

A similar circumstance was realized by Deshpande and He \cite{DH} to apply
to two cases: (a) An isospin triangle for $B \to \pi \pi$ decays, while
continuing to hold, receives small contributions from electroweak penguins.
This can in principle affect the analysis proposed in \cite{GL} for
extracting the weak phase $\alpha$. (b) The validity of the SU(3) triangle
proposed in \cite{BPP,PRL,PLB}, involving the comparison of $B \to \pi \pi$
and $B \to \pi K$ decays, is also affected.

\begin{figure}
\centerline{\psfig{figure=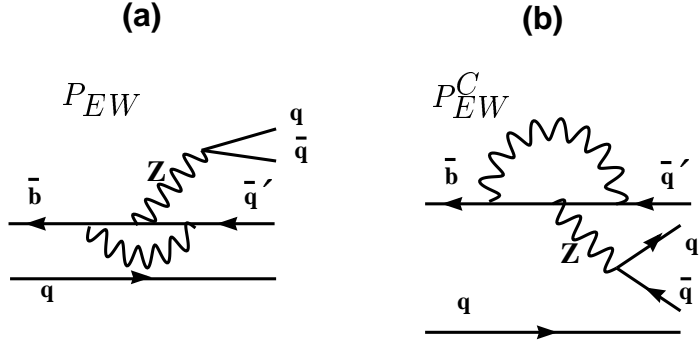,height=7.0cm}}
\caption{(a) Color-allowed $Z$-penguin, (b) Color-suppressed $Z$-penguin.}
\label{Zpenguin}
\end{figure}

The dominant electroweak penguin contribution arises from $Z$ exchange.
There are two such diagrams, shown in Fig.~\ref{Zpenguin}. The distinction
between the two is that the diagram of Fig.~\ref{Zpenguin}(a) is
color-allowed, while that of Fig.~\ref{Zpenguin}(b) is color-suppressed.
We refer to these as $\pew$ and $\pewc$, respectively.
Thus, EWP effects will be most important when the $\pew$ diagram is
involved, that is, when there is a nonstrange neutral particle in the final
state, such as $\pi^0$, $\eta$, $\rho^0$ or $\phi$. All-charged final
states will be less affected by the presence of electroweak penguins, since
in this case only the $\pewc$ diagram can arise.  EWP diagrams which involve
the annihilation of the quarks in the initial $B$ meson are suppressed by a
factor of $f_B/m_B \approx 5\%$. As we will see from the hierarchy of
diagrams discussed in the next section, this means that we will always be
able to ignore annihilation-type EWP diagrams.

The ratio of a $\pew$ electroweak penguin to a gluonic penguin contribution
$P$ in $b$ quark decays contains a factor of $\alpha_2/\alpha_s \approx
(1/30)/0.2 \approx 1/6$, where we have evaluated both couplings at $m_b$.
The electroweak penguin for $Z$ exchange contains a factor of $(m_t/M_Z)^2
\approx 4$ in contrast to a logarithm $\ln(m_t^2/m_c^2) \approx 9$ in the
gluonic penguin. Thus, the overall electroweak penguin's amplitude should
be ${\cal O}(10\%)$ that of the gluonic penguin, modulo group-theoretic
factors. This is in qualitative accord with the result of \cite{DH}.

A more quantitative calculation of the ratio $\pew/P$ will necessarily
involve hadronic physics. In particular, the matrix elements for $\pew$ and
$P$ are almost certainly different, since the two diagrams clearly have
different dynamical structures. Such model-dependent calculations are
fraught with uncertainties \cite{uncertainties}. (For example, although it
might be argued that factorization applies to the $\pew$ diagram, it is
considerably more doubtful for $P$.) Thus, theoretical calculations of
$\pew/P$ \cite{DH} should be viewed with a certain amount of skepticism.
Still, the magnitude of this ratio is very important. As we will see in the
following sections, the methods presented in \cite{BPP,PRL,PLB} for the
extraction of weak and strong phases will be invalidated if EWP's are too
large, say $\pew/P \gsim 20\%$. For these reasons it is important to try to
obtain information about electroweak penguins from the data.

\bigskip
\leftline{\bf B. Diagrams and hierarchies}
\bigskip

There are, of course, other diagrams which contribute to $B\to PP$ decays,
and it is equally important to estimate the size of electroweak penguins
relative to these other contributions.

Excluding electroweak penguins, there are six distinct diagrams which
contribute to $B$ decays: (1) a (color-favored) ``tree'' amplitude $T$,
$T'$; (2) a ``color-suppressed'' amplitude $C$, $C'$; (3) a ``penguin''
amplitude $P$, $P'$; (4) an ``exchange'' amplitude $E$, $E'$; (5) an
``annihilation'' amplitude $A$, $A'$; (6) a ``penguin annihilation''
amplitude $PA$, $PA'$.  (We refer the reader to Ref.~\cite{BPP}
or~\cite{GHLR} for a more complete discussion of the diagrams.) For $T$,
$C$, $E$ and $A$, the unprimed and primed amplitudes contribute to the
decays $\bar b \to \bar u u \bar d$ and $\bar b \to \bar u u \bar s$,
respectively, and the primed amplitudes are related to their unprimed
counterparts by a factor of $|V_{us}/V_{ud}| \simeq \lambda = 0.22$. For
$P$ and $PA$ the unprimed and primed amplitudes contribute to the decays
$\bar b \to \bar d$ and $\bar b \to \bar s$, respectively. In this case,
the primed amplitudes are actually {\it larger} than the unprimed
amplitudes by a factor of $|V_{ts}/V_{td}|$, which is of order $1/\lambda$.

In Ref.~\cite{GHLR} we estimated the relative sizes of these diagrams
in $B\to PP$ decays. Here we include electroweak penguins, justifying
our estimates of their magnitudes after presenting the expected
hierarchies.
\smallskip

\noindent
{\it 1. $\bar b \to \bar u u \bar d$ and $\bar b \to \bar d$ transitions:}
The dominant diagram is $T$. Relative to the dominant contribution, we
expect
\begin{eqnarray}
\label{buudhierarchy}
1 & : & |T|, \nonumber \\
{\cal O}(\lambda) & : & |C|,~|P|, \nonumber \\
{\cal O}(\lambda^2) & : & |E|,~|A|,~|\pew| \nonumber \\
{\cal O}(\lambda^3) & : & |PA|,~|P_{EW}^{\prime C}|.
\end{eqnarray}
\smallskip

\noindent
{\it 2. $\bar b \to \bar u u \bar s$ and $\bar b \to \bar s$ transitions:}
Here the dominant diagram is $P'$. Relative to this, we estimate
\begin{eqnarray}
\label{buushierarchy}
1 & : & |P'|, \nonumber \\
{\cal O}(\lambda) & : & |T'|,~|\pew'|\nonumber \\
{\cal O}(\lambda^2) & : & |C'|,~|PA'|,~|P_{EW}^{\prime C}| \nonumber \\
{\cal O}(\lambda^3) & : & |E'|,~|A'|.
\end{eqnarray}
The use of the parameter $\lambda = 0.22$ here is unrelated to CKM matrix
elements -- it is simply used as a measure of the approximate relative
sizes of the various contributions. For instance, $|C/T| \sim \lambda$ is
due to color suppression, while $E$ and $A$ are suppressed relative to $T$
by the factor $f_B/m_B \approx 0.05 \sim \lambda^2$. Similarly, $PA/P \sim
f_B/m_B$. Although it is fairly certain that $P'$ dominates the second
class of decays, the value of the ratio $|T'/P'|$ is less clear. Our value
of $\lambda$ for this ratio is probably a reasonable estimate. Finally as
discussed in Ref.~\cite{GHLR}, we expect the SU(3) corrections to a diagram
to be roughly 20\% ($\sim \lambda$) of that particular diagram. We shall
discuss SU(3)-breaking effects in the cases of several specific processes
of interest in Sections III and IV.

Note that both of the above hierarchies are educated guesses -- it is
important not to take them too literally. Since $\lambda$ is not that small
a number, a modest enhancement or suppression (due to hadronic matrix
elements, for example) can turn an effect of ${\cal O}(\lambda^n)$ into an
effect of ${\cal O}(\lambda^{n\pm 1})$. Ultimately experiment will tell us
exactly how large the various diagrams are.

Some combination of the decays $B^0 \to \pi^+\pi^-$ and $B^0 \to K^+\pi^-$
has been observed \cite{Kpisep}. The most likely branching ratios for these
two modes are both about $10^{-5}$ (though all that can be conclusively
said is that their sum is about $2 \times 10^{-5}$). One then concludes
that the $T$ and $P'$ amplitudes are about the same size. In this case, the
estimated hierarchies in Eqs.~(\ref{buudhierarchy}) and
(\ref{buushierarchy}) can be combined.

The above estimated hierarchies can be used to judge how large electroweak
penguin effects should be. Our naive estimate of $\pew/P$ was ${\cal
O}(10\%)$. Allowing for some variation in either direction, we have $\pew/P
\sim {\cal O}(\lambda)$ -- ${\cal O}(\lambda^2)$. Thus, for $\bar b \to
\bar u u \bar d$/$\bar b \to \bar d$ decays, EWP's are at most ${\cal
O}(\lambda^2)$ of the dominant $T$ contribution. For this reason it is
unlikely that electroweak penguins will significantly affect $B\to\pi\pi$
decays. On the other hand, for $\bar b \to \bar u u \bar s$/$\bar b \to
\bar s$ decays, EWP contributions can be as much as ${\cal O}(\lambda)$ of
the dominant $P'$ diagram, which is why they may be important in $B\to\pi
K$ decays.

As discussed in the previous section the color-suppressed electroweak
penguin $\pewc$ should be smaller than its color-allowed counterpart $\pew$
by approximately a factor of $\lambda$. Thus this contribution is probably
completely negligible in $\bar b \to \bar u u \bar d$/$\bar b \to \bar d$
decays, and is at most a 5\% effect in $\bar b \to \bar u u \bar s$/$\bar b
\to \bar s$ decays relative to the dominant $P'$ contribution.

\bigskip
\centerline{\bf III. $B\to PP$ DECAYS}
\bigskip

\leftline{\bf A. Decomposition in terms of SU(3) amplitudes}
\bigskip

We review briefly the SU(3) discussion of \cite{BPP}. The weak Hamiltonian
operators associated with the transitions $\bar b \to \bar q u \bar u$ and
$\bar b \to \bar q$ ($q = d$ or $s$) transform as a ${\bf 3^*}$, ${\bf 6}$,
or ${\bf 15^*}$ of SU(3). These combine with the triplet light quark in the
$B$ meson and couple to a symmetric product of two octets (the pseudoscalar
mesons) in the final state, leading to decays characterized by one singlet,
three octets, and one ${\bf 27}$-plet amplitude. Separate amplitudes apply
to the cases of strangeness-preserving and strangeness-changing
transitions. The diagrams $T$--$PA$ are a useful representation of flavor
SU(3) amplitudes. Although there are 6 types of diagram (excluding
electroweak penguins), they only appear in 5 linear combinations in $B\to
PP$ decays, in accord with the group theory result.

The inclusion of electroweak penguins does not affect this picture. The
ratio of transitions $\bar b \to \bar q u \bar u$, $\bar b \to \bar q d
\bar d$, and $\bar b \to \bar q s \bar s$ is altered, but the $\bar b \to
\bar q d \bar d$ and $\bar b \to \bar q s \bar s$ terms remain equal.
(This is obvious for the $\gamma$- and $Z$-penguins. For the box diagrams,
this equality is ensured by the GIM mechanism. There are contributions from
the boxes which break this equality, but they are much suppressed relative
to the dominant term.) The weak Hamiltonian thus continues to contain terms
transforming as a ${\bf 3^*}$, ${\bf 6}$, or ${\bf 15^*}$ of SU(3), but in
different proportions. Thus, even if one includes electroweak penguin
graphs, there must continue to be five independent amplitudes describing
$\Delta S = 0$ decays and five other amplitudes describing $|\Delta S| = 1$
decays. However, some of the correspondence between $\Delta S = 0$ and
$|\Delta S| = 1$ decays present in the previous description will be
altered. In this section we extend the decomposition of $B\to PP$ decays in
terms of the diagrams $T$--$PA$ to include the electroweak penguin diagrams
of Fig.~\ref{Zpenguin}. In this way we see explicitly how $B\to\pi\pi$ and
$B\to\pi K$ decays are affected by electroweak penguins.

In \cite{BPP} it was argued that the diagrams $E$, $A$ and $PA$ (and their
primed counterparts) are negligible since they are suppressed by a factor
of $f_B/m_B = {\cal O} (\lambda^2)$ and hence are unlikely to be important
in many cases.  However, there are processes such as $B^0\to \pi^0\pi^0$,
$B^+\to K^+\bar K^0$ and $B_s\to \pi^0\bar K^0$ which are dominated by the
$\olambdai$ terms $C$ and/or $P$.  In these cases diagrams suppressed by
$\olambda 2$ with respect to the dominant $T$ contributions, such as $E,A$
and $\pew$, can cause a significant change in the rate. There are
situations, which we will soon discuss, when one cannot neglect such
seemingly small diagrams. These are precisely the cases where EWP's are
important.

We continue to use the approximation of ignoring $E$, $A$ and $PA$-type
diagrams when considering electroweak penguin effects as long as their
effects are $\olambda2$ with respect to the dominant contribution to a
process. Annihilation-type electroweak penguin amplitudes will always be
subdominant by at least $\olambda2$ in all the processes we will consider
and hence we can ignore them. In $|\Delta S = 1|$ decays, the $C'$
contribution should really be dropped, since it is expected to be of the
same order as the $PA'$ diagram, which has been neglected. Nevertheless, we
continue to keep track of the $C'$ contribution in such decays, since it is
related to the non-negligible $C$ diagram in $\Delta S = 0$ decays.
(Obviously our results should not, and do not, depend on keeping or
ignoring the $C'$ contribution.)

The distinction between the gluonic penguin $P$ and the electroweak penguin
$\pew$ is the coupling to the light quarks. In $P$, the quarks $u$, $d$ and
$s$ have equal couplings to the gluon. In $\pew$, however, the $u$ and
$d$/$s$ quarks are treated differently. Schematically, we can represent the
couplings of the strong and electroweak penguins as follows:
\begin{eqnarray}
\label{cucd}
P & : & u{\bar u} + d{\bar d} + s{\bar s} ~~~, \nonumber \\
\pew,~\pewc & : & c_u \, u{\bar u} + c_d (d{\bar d} + s{\bar s})~~~.
\end{eqnarray}
Although the precise values of $c_u$ and $c_d$ depend on the detailed
structure of the electroweak penguin, they are taken to be numbers of order
1. For example, if the electroweak penguin coupled to the charge of the
quarks (as it would if it arose purely from photon exchange), we would have
$c_u=2/3$ and $c_d=-1/3$.

In Tables \ref{BPPtablei} and \ref{BPPtableii} we present the decomposition
of the 13 $B\to PP$ decays in terms of the various diagrams, for $P = \pi$
or $K$.  We warn the reader that non-negligible SU(3)-breaking corrections
can lead to differences in certain decays that appear equal in the above
Tables. For example, according to Table 2, $B^+\to\pi^+K^0$ and $B_s \to
K^0 \bar K^0$ will have the same rate. However, SU(3)-breaking effects
introduce a rate difference here. We refer the reader to Ref.~\cite{GHLR}
for more details. We shall, however, correctly include SU(3)-breaking
effects when discussing specific examples in the following sections.

\bigskip
\leftline{\bf B. Effects on CP analyses}
\bigskip

There are several interesting aspects of Tables \ref{BPPtablei} and
\ref{BPPtableii} worth mentioning.
\smallskip

\begin{table}
\caption{Decomposition of $B\to PP$ amplitudes for $\Delta C= \Delta S=0$
transitions in terms of graphical contributions of
Refs.~\protect\cite{BPP},~\protect\cite{GHLR} and Fig.~1. For completeness
we include color-suppressed $\pewc$ contributions even when they are
estimated to be negligible.}
\begin{center}
\begin{tabular}{l l c c c c c c} \hline
           &  Final        &  $T$,$C$,$P$        &Electroweak \\
           &  state        & contributions  & Penguins    \\ \hline
$B^+\to$ & $\pi^+\pi^0$ & ${ -(T+C)/\s}$
                                & $-[ (c_u-c_d)\pew +(c_u-c_d)\pewc]/\s$  \\
         & $K^+ \bar K^0$ & $P+A$ & $c_d\pewc$   \\ \hline
$B^0\to$ & $\pi^+\pi^-$ & ${ -(T+P)} $
                                &$- c_u\pewc $  \\
         & $\pi^0\pi^0$ & ${ -(C-P-E)/\s}$
                                &$-[{ (c_u-c_d)\pew} + c_d\pewc]/\s$  \\
         & $K^0 \bar K^0$ & ${ P}$
                                &$ c_d\pewc$   \\ \hline
$B_s\to$ & $\pi^+K^-$ & ${ -(T+P)}$
                                & $-c_u\pewc$ \\
         & $\pi^0 \bar K^0$ & ${ -(C-P)/\s}$
                        &$-[{ (c_u-c_d)\pew} - c_d\pewc]/\s$  \\ \hline
\end{tabular}
\end{center}
\label{BPPtablei}
\end{table}

\begin{table}
\caption{Decomposition of $B \to PP$ amplitudes for $\Delta C = 0,~
|\Delta S| = 1$  transitions in terms of graphical contributions of
Ref.~\protect\cite{BPP},~\protect\cite{GHLR} and Fig.~1. For completeness
we include $C'$ and the color-suppressed $P_{EW}^{\prime C}$
contributions  even though they are estimated to be negligible. }
\begin{center}
\begin{tabular}{l l c c c c c c} \hline
           &  Final        & $P'$,$T'$,$C'$    &Electroweak \\
           &  state        & contributions  & Penguins    \\ \hline
$B^+\to$ & $\pi^+K^0$ & ${ P'}$
                                        &$ c_d P_{EW}^{\prime C}$ \\
         & $\pi^0K^+$ & $-({ P'+T'}+C')/\s$
                &$-[{ (c_u-c_d)\pew'}+c_u P_{EW}^{\prime C}]/\s$ \\ \hline
$B^0\to$ & $\pi^-K^+$ & ${ -(P'+T')}$
                                        &$- c_u P_{EW}^{\prime C}$  \\
         & $\pi^0K^0$ & $-({ P'}-C')/\s$
                &$-[{ (c_u-c_d)\pew'} - c_d P_{EW}^{\prime C}]/\s$  \\ \hline

$B_s\to$ & $K^+K^-$ & ${ -(P'+T')}$
                                        &$-c_u P_{EW}^{\prime C}$  \\
	 & $K^0 \bar K^0$ & ${ P'}$
                                        &$c_d P_{EW}^{\prime C}$  \\  \hline
\end{tabular}
\end{center}
\label{BPPtableii}
\end{table}

\noindent
{\it 1. $B\to \pi\pi$ decays:}

Consider the $B\to\pi\pi$ decays in Table \ref{BPPtablei}. The decay
$B^+\to\pi^+\pi^0$, which is purely $I=2$, has an electroweak penguin
component. If our estimated hierarchy is accurate, this component should be
between ${\cal O}(\lambda^2)$ and ${\cal O}(\lambda^3)$ of the dominant $T$
contribution. This is in agreement with Deshpande and He \cite{DH}, who
find that $|A_{EWP}/A_T| \approx 1.6\%\, |V_{td}/V_{ub}|$ for this decay.
In other words, the EWP contribution to $B^+\to\pi^+\pi^0$ is very small.
It is even smaller in the decay $B^0\to\pi^+\pi^-$, since only the
color-suppressed EWP can contribute here. On the other hand, electroweak
penguins can be more significant in $B^0\to\pi^0\pi^0$ decays, since this
decay suffers color suppression.

\begin{figure}
\centerline{\psfig{figure=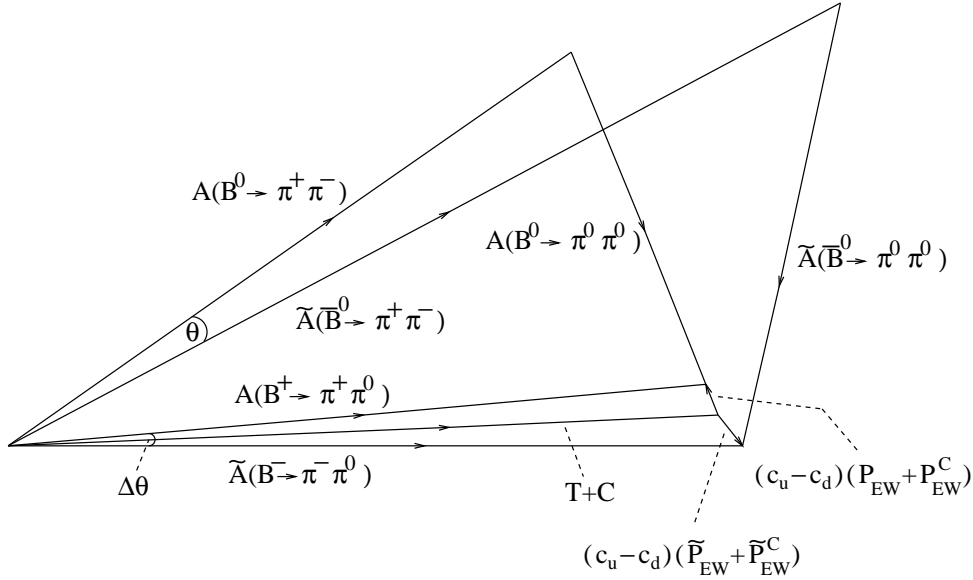,angle=90,height=3.0in}}
\caption{Isospin analysis of $B\to \pi\pi$ decays with the inclusion of
electroweak penguins. The amplitudes $\tilde{A}(\bar B\to\pi\pi)$ are defined
as ${\rm exp}(2i\gamma)A(\bar B\to\pi\pi)$, with similar definitions for
$\tilde{P}_{EW}$ and $\tilde{P}_{EW}^C$.}
\label{isoanalysis}
\end{figure}

The size of EWP's is relevant to the extraction of $\alpha$ via the
analysis proposed in~\cite{GL}. Let us study this effect in detail. This
analysis requires measuring the (time-integrated) rates of
$B^+\to\pi^+\pi^0$, $B^0\to\pi^+\pi^-$, $B^0\to\pi^0\pi^0$ and their
CP-conjugate counterparts, and observing the time-dependence of
$B^0(t)\to\pi^+\pi^-$. The amplitudes of these six processes form two
triangles, as shown in Fig.~\ref{isoanalysis}, in which the CP-conjugate
amplitudes have been rotated by a common phase $\tilde{A}(\bar
B\to\pi\pi)\equiv {\rm exp}(2i\gamma)A(\bar B\to\pi\pi)$
(and similarly for $\tilde{P}_{EW}$ and $\tilde{P}_{EW}^C$).
The CKM phase $\alpha$ is measured from the time-dependent rate of
$B^0(t)\to\pi^+\pi^-$, which involves a term
\beq
\left\vert {\tilde{A}(\bar{B}^0\to\pi^+\pi^-)\over A(B^0\to\pi^+\pi^-)}
\right\vert \sin(2\alpha+\theta)\sin(\Delta mt)~,
\eeq
where $\Delta m$ is the neutral $B$ mass difference. The angle $\theta$
is measured as shown in Fig.~\ref{isoanalysis}.

The effect of the EWP amplitudes on determining $\theta$ and
correspondingly fixing $\alpha$ is rather clearly represented by the small
vectors at the right bottom corner of the Fig.~\ref{isoanalysis}. These
terms, given by $(c_u-c_d)(P_{EW}+P_{EW}^C)$ and its CP-conjugate, have
unknown phases relative to the $T+C$ term which dominates
$A(B^+\to\pi^+\pi^0)$ and its charge-conjugate. This leads to a very small
uncertainty in the relative orientation of the two triangles. [In the limit
of neglecting EWP amplitudes, one would have
$\tilde{A}(B^-\to\pi^-\pi^0)=A(B^+\to\pi^+\pi^0)$]. The uncertainty in
measuring $\theta$, and consequently in determining $\alpha$, is given by
\beq
\Delta\alpha\approx{1\over 2}\Delta\theta \leq \left\vert
{(c_u-c_d)(P_{EW}+ P_{EW}^C)\over T+C} \right\vert~.
\eeq
We therefore conclude that the effects of EWP amplitudes on the measurement
of $\alpha$ are at most of order $\lambda^2$ and are negligible.

Since a different conclusion has been claimed in~\cite{DH,Desh}, let us clarify
the apparent disagreement. The authors of~\cite{DH,Desh} have only shown that
the error in determining $\alpha$ from the rate of $B^0(t)\to\pi^+\pi^-$ is
large. This is dominantly the effect of the gluonic penguin, as already noted
in~\cite{MG}. They have not separated the effect of EWP amplitudes.
Fig.~\ref{isoanalysis} shows clearly how small this effect is.
\smallskip

\noindent
{\it 2. $B\to \pi K$ decays:}

We now turn to the $B\to\pi K$ decays in Table~\ref{BPPtableii}. In the
absence of electroweak penguins, one can write two triangle relations
involving amplitudes in both the $\Delta S = 0$ and $|\Delta S| = 1$
sectors:
\begin{eqnarray}
\label{tri-rel}
\s A(B^0 \to \pi^0 K^0) + A(B^0 \to \pi^- K^+) &=& \lambda \s
A(B^+ \to \pi^+ \pi^0) \nonumber \\
- (C'-P') - (P'+T') & = & - \lambda (T + C)
\end{eqnarray}
\begin{eqnarray}
\label{PRLtriangle}
\s A(B^+\to\pi^0 K^+) + A(B^+\to\pi^+ K^0) & = & \lambda \s
A(B^+\to\pi^+\pi^0) \nonumber \\
- (T'+C'+P') + (P') & = & - \lambda (T + C)
\end{eqnarray}
SU(3) breaking can be taken into account by including a factor of
$f_K/f_\pi$ on the right-hand side \cite{GHLR}. In Eq.~(\ref{PRLtriangle})
above, SU(3) relates the $I=3/2$ $\pi K$ amplitude to the $I=2$ $\pi\pi$
amplitude. By measuring the three rates involved in the triangle relation,
as well as their CP-conjugates, the weak CKM angle $\gamma= {\rm
Arg}(V_{ub}^*)$, which is the weak phase of $A(B^+\to\pi^+\pi^0)$, can be
extracted \cite{PRL}. By using both Eqs.~(\ref{tri-rel}) and
(\ref{PRLtriangle}), strong final-state phases and the sizes of the
different diagrams can also be extracted \cite{PLB}.

When electroweak penguins are included, however, these two triangle
relations no longer hold. For example, the left-hand side of
Eq.~(\ref{PRLtriangle}) is now equal to
\beq
-[T'+C'+(c_u-c_d)(\pew'+P_{EW}^{\prime C})],
\eeq
while the right-hand side is
\beq
- \lambda [T+C+(c_u-c_d)(\pew+\pewc)].
\eeq
Despite their similarity, these two expressions are not equal since the
relation between non-penguin contributions ($T'/T = C'/C = \lambda$) does not
hold for the electroweak penguins: $|\pew'/\pew| = |V_{ts}/V_{td}| \sim
1/\lambda$. This relation would only hold if $c_u$ were equal to $c_d$, which
cannot happen since EWP's are not isosinglets.

{}From our previous discussion, we estimate that $|\pew'/T'|$
may be as much as $\sim 1$. Eventually, it will be up to experiment to
determine the size of electroweak penguins. However, in a realistic
scenario, with hierarchies such as those discussed Sec.~II B, EWP's lead to
large uncertainties in the extraction of weak CKM angles and strong phases
through the analyses of Refs.~\cite{PRL,PLB}. In Sec. IV we extend the
SU(3) triangle analysis of Ref.~\cite{PRL} to a quadrangle relation, using
more decay rate measurements to exhibit a new way of measuring the weak
angle $\gamma$ which holds even in the presence of electroweak penguins.

\newpage
\leftline{\bf C. Experimental signals}
\bigskip

As discussed above, the fate of the analyses of Refs.~\cite{PRL,PLB} for
extracting weak CKM phase information depends crucially on the size of
electroweak penguins. Rather than relying on theoretical calculations, which
inevitably have uncertainties due to hadronic matrix elements, it would be
preferable to obtain this information from experiment.

Electroweak penguins are expected to dominate decays of the form $B_s \to
(\phi~{\rm or}~\eta) + (\pi~{\rm or}~\rho)$ \cite{DHT}. This is easy to
understand in terms of diagrams:
\beq
\label{ewpdominant}
A[B_s \to (\phi~{\rm or}~\eta) + (\pi~{\rm or}~\rho)] \sim
-C' + E' - (c_u-c_d) \pew' ~.
\eeq
We have already argued that the $E'$ diagram is small, so, from
Eq.~(\ref{buushierarchy}) and the discussion following it, we see that the
dominant contribution is $\pew'$.

Unfortunately, even though these decays are dominated by electroweak
penguins, their branching ratios are all small, less than $O(10^{-6})$.
Furthermore, they all involve the decays of $B_s$ mesons, which are not as
accessible experimentally. This leads to the obvious question: are there
signals for electroweak penguins which involve decays of $B^\pm$ or $B^0$
mesons, and which have large branching ratios? Indeed there are. Consider
the decays $B^+ \to \pi^0 K^+$ and $B^0\to\pi^- K^+$. From Table
\ref{BPPtableii}, we have
\beq
\s A(B^+ \to \pi^0 K^+) \simeq -[T'+P'+(c_u-c_d) \pew' ] ~~,~~~
A(B^0\to\pi^- K^+) \simeq -[T'+P'] ~~~,
\eeq
where we have dropped the (much smaller) terms $C'$ and $P_{EW}^{\prime C}$.
Both of these decays should have branching ratios of ${\cal O}(10^{-5})$ as a
result of the dominant $P'$ contribution. A difference in the branching ratios
of these decays can only be due to the presence of electroweak penguins. Though
indirect, this is very likely to be the first experimental test of such
effects. Similarly, the most likely source of a difference in the branching
ratios of $B^0\to\pi^0 K^0$ and $B^+ \to \pi^+ K^0$ will be the contribution of
electroweak penguins.

\bigskip
\centerline{\bf IV. AMPLITUDE QUADRANGLES}
\bigskip

\leftline{\bf A. SU(3)-invariant analysis for $B \to \pi K$}
\bigskip

The decays $B \to \pi K$ involve a weak Hamiltonian with both $I = 0$ and
$I = 1$ terms. The $I=0$ piece can lead only to a $\pi K$ final state with
$I = 1/2$, while the $I = 1$ piece can lead to both $I = 1/2$ and $I = 3/2$
final states. Thus, there are two decay amplitudes leading to $I_{\pi K} =
1/2$ and one leading to $I_{\pi K} = 3/2$. Since there are four amplitudes
for $B \to \pi K$ decays, they satisfy a quadrangle, which we may write as
\cite{NQ,GQ}
\beq
A(B^+ \to \pi^+ K^0) + \sqrt{2} A(B^+ \to \pi^0 K^+) =
\sqrt{2} A(B^0 \to \pi^0 K^0) + A(B^0 \to \pi^- K^+) = A_{3/2}~~~.
\label{eqn:quad}
\eeq

\begin{figure}
\centerline{\psfig{figure=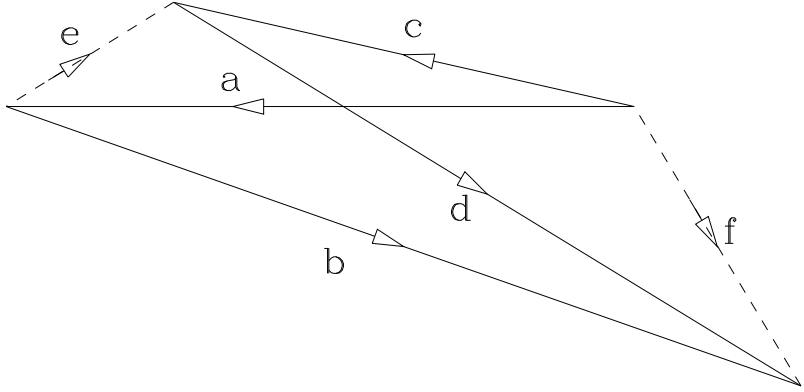,height=3in}}
\caption{Amplitude quadrangle for $B \to \pi K$ decays. (a) $A(B^+ \to
\pi^+ K^0)$; (b) $\protect \sqrt 2 A(B^+ \to \pi^0 K^+)$; (c) $\protect
\sqrt 2 A(B^0 \to \pi^0 K^0)$; (d) $A(B^0 \to \pi^- K^+)$; (e) the diagonal
$D_2 = \protect\sqrt 3 A(B_s \to \pi^0 \eta)$; (f) the diagonal $D_1 =
A_{3/2}$ corresponding to the $I = 3/2$ amplitude.}
\label{piKquad}
\end{figure}

With the phase conventions adopted in \cite{BPP}, the quadrangle has the
shape shown in Fig.~\ref{piKquad}, with two short diagonals. These
diagonals are:
\beqn
D_1 & = & -[T'+C'+(c_u-c_d)(\pew'+P_{EW}^{\prime C})] ~~~, \nonumber\\
D_2 & = & -C' -(c_u-c_d) \pew' - A' ~~~.
\eeqn
The first of these diagonals, $D_1$, is just the amplitude $A_{3/2}$. The
key point is that $A(B_s \to \pi^0 \eta) = -[C' +(c_u-c_d) \pew' -
E']/\sqrt{3}$, for an octet $\eta$. Thus, ignoring the very small $E'$ and
$A'$ diagrams, the second diagonal, $D_2$, is
in fact equal to $\sqrt{3} A(B_s \to \pi^0 \eta)$. Therefore the shape of
the quadrangle is uniquely determined, up to possible discrete ambiguities.
The case of octet-singlet mixtures in the $\eta$ simply requires us to
replace the $\sqrt{3}$ by the appropriate coefficient \cite{GK}, since one
can show that the singlet piece of $\eta$ does not contribute appreciably
here.

The quadrangle has been written in such a way as to illustrate the fact,
noted in Refs.~\cite{BPP,PRL,PLB}, that the $B^+ \to \pi^+ K^0$ amplitude
receives only penguin contributions in the absence of ${\cal O}(f_B/m_B)$
corrections. The weak phases of both gluonic and electroweak $\bar b \to
\bar s$ penguins, which are dominated by a top quark in the loop, are
expected to be $\pi$. We have oriented the quadrangle to subtract out the
corresponding strong phase.

The $I = 3/2$ amplitude is composed of two parts:
\beq
A_{3/2} = |A^T_{\pi K}| e^{i \gamma} e^{i \tilde \delta_T} - |A^{\rm
EWP}_{\pi K}| e^{i \tilde \delta_{\rm EWP}}~~~,
\eeq
where we have explicitly exhibited electroweak and final-state phases, and
the tildes denote differences with respect to the strong phase shift in the
$B^+ \to \pi^+ K^0$ amplitude. The corresponding charge-conjugate
quadrangle has one diagonal equal to
\beq
\bar A_{3/2} = |A^T_{\pi K}| e^{- i \gamma} e^{i \tilde \delta_T} - |A^{\rm
EWP}_{\pi K}| e^{i \tilde \delta_{\rm EWP}}~~~,
\eeq
so that one can take the difference to eliminate the electroweak penguin
contribution:
\beq \label{eqn:diff}
A_{3/2} - \bar A_{3/2} = |A^T_{\pi K}| 2 i \sin \gamma e^{i \tilde
\delta_T} ~~~.
\eeq
In diagrammatic language, the quantity $|A^T_{\pi K}|$ is just $|T'+C'|$.
But this can be related to the $I = 2$ $\pi \pi$ amplitude in order to
obtain $\sin  \gamma$. Specifically, if we neglect electroweak penguin
effects in $B^+ \to \pi^+ \pi^0$ (a good approximation, as noted in
Sec.~IIIB), we find that
\beq \label{eqn:pipi}
|A^T_{\pi K}| = \lambda (f_K/f_\pi) \sqrt{2} |A(B^+ \to \pi^+ \pi^0)|~~~.
\eeq
Thus, we can extract not only $\sin \gamma$, but also a strong phase shift
difference $\tilde{\delta}_T$, by comparing (\ref{eqn:diff}) and
(\ref{eqn:pipi}). Of course, if such a strong phase shift difference
exists, the $B$ and $\bar B$ quadrangles will necessarily have different
shapes, and CP violation in the $B$ system will already have been
demonstrated.

We should remark that the quadrangle construction for $B \to \pi K$ decays
introduced in \cite{NQ} and refined in \cite{GQ} assumed the presence of a
single weak phase in the amplitude $A_{3/2}$, and no longer is valid in the
presence of electroweak penguins.

\bigskip
\leftline{\bf B. SU(3)-breaking effects in $B \to \pi K$}
\bigskip

The analysis presented above relies on the equality of two small amplitudes
-- the diagonal $D_2$ of the $\pi K$ quadrangle and the decay amplitude
$\sqrt{3} A(B_s \to \pi^0 \eta)$. Thus one might worry that small effects,
which we have ignored up to now, might break this equality. We address this
question here.

First, we have ignored $E'$ and $A'$ diagrams in equating these two
amplitudes. This should not cause any problems. We expect that $\pew'$ is
roughly of the same size as $T'$. But $E'$ and $A'$ are suppressed by
$f_B/m_B \approx 5\%$ relative to $T'$. Thus their neglect introduces at
most a small error into our analysis.

The second possibility involves SU(3) breaking. The effects of SU(3)
breaking in two-body decays of $B$ mesons have been analyzed by us in more
detail in a longer paper \cite{GHLR}. The largest terms in the present case
involve the effect of SU(3) breaking on the dominant gluonic penguin term
($P'$) in $B \to K \pi$. These terms are of the same strength in all the $B
\to K \pi$ amplitudes illustrated in Fig.~\ref{piKquad}, and hence cancel
in the construction of the two diagonals. The next most important term
involves SU(3) breaking in the ratio of the $|\Delta S| = 1$ and $\Delta S
= 0$ non-penguin amplitudes. However, this is expected to be
well-approximated by the ratio
$f_K/f_{\pi}$ \cite{GHLR} (see also
\cite{BPP,SilWo}), as in Eq.~(\ref{eqn:pipi}). The critical term turns out
to be the effect of SU(3) breaking on the electroweak penguin.
Specifically, the $B_s \to \pi^0 \eta$ decay involves a spectator $s$
quark, whereas the spectator quark in the $B\to\pi K$ decays is $u$ or $d$.
Thus, the SU(3) breaking corresponds here to a difference in the form
factors for the two types of decays. Although we expect SU(3)-breaking
effects to be typically of order 25\% (i.e.\ the difference between $f_\pi$
and $f_K$), here they
are expected to be smaller, since the mass ratio $m_\eta/m_K$ is much
closer to unity than is $m_K/m_\pi$. Still, this SU(3) breaking does
introduce some theoretical uncertainty into this method for obtaining
$\gamma$.

\bigskip
\leftline{\bf C. The processes $B \to \pi K^*$ and $B \to \rho K$}
\bigskip

We have carried out a similar analysis for the decays $B \to \pi K^*$.
Clearly it is still possible to write an amplitude quadrangle for
these processes; the question is simply the interpretation of the
diagonals.

There are more SU(3) amplitudes in $B\to PV$ decays since the final-state
particles do not belong to the same octet. Nevertheless, one can still use
a graphical analysis in the spirit of Ref.~\cite{BPP} -- there are just
more diagrams. For example, instead of one $T$ diagram, there are two
($T_P$ and $T_V$), corresponding to the cases where the spectator quark
hadronizes into the $P$- or $V$-meson in the final state.

Carrying out such a graphical analysis, we find that the diagonals of the
$\pi K^*$ quadrangle are
\beqn
D_1^* & = & -[T_P'+C_P'+(c_u-c_d) (P_{EW,V}' + P_{EW,V}^{\prime C})] ~~~,
\nonumber\\
D_2^* & = & -C_P' -(c_u-c_d) P_{EW,V}' ~~~,
\eeqn
where the subscripts $P$ and $V$ represent the spectator quark hadronizing
into the $\pi$ and $K^*$, respectively. (In the above we have ignored
annihilation-type contributions.) Remarkably, the diagonal $D_2^*$ (labeled
by (e) in Fig.~\ref{piKquad}) corresponds to $\sqrt{2} A(B_s \to \pi^0
\phi)$. Again, the shape of the quadrangle can be specified by experimental
measurements! The other diagonal $D_1^*$ contains both an electroweak
penguin piece (which we can eliminate in the manner noted in Sec.~IV A
above), and a non-penguin piece $-(T_P' + C_P')$. This latter piece is
closely related to the amplitude for the decay $B^+ \to \pi^0 \rho^+$:
\beq
\s A(B^+ \to \pi^0 \rho^+) = -[T_P + C_P - P_P + P_V + (c_u-c_d)
P_{EW,V}]~.
\label{pirhoamp}
\eeq
If the penguin diagrams are unimportant in this decay, or if the two types
of penguin contributions $P_P$ and $P_V$ cancel (the EWP is expected to be
quite small here), the analysis can be carried through exactly as in
Sec.~IV A.
In this case, the precision on the measurement of $\gamma$ is roughly of
order $|(P_P-P_V)/(T_P+C_P)|$.

Another quadrangle relation holds for the amplitudes of $B\to \rho K$. They
are obtained from the amplitudes of $B \to \pi K^*$ by replacing $T_P',
C_P', P_{EW,V}'$, etc., by $T_V', C_V', P_{EW,P}'$, etc. Here one of the
diagonals of the quadrangle is given by $\sqrt{3}A(B_s \to \eta \rho^0)$.
The other diagonal (obtained from $D_1^*$ by substituting $P
\leftrightarrow V$ in (18)) contains $-(T_V'+C_V')$ and an electroweak
penguin term. When the latter is eliminated as in Sec.~IV A, the remaining
$-(T_V'+C_V')$ term is approximately equal to $\sqrt{2}A(B^+ \to \pi^+
\rho^0)$.

\bigskip
\centerline{\bf V. DATA: STATUS AND PROSPECTS}
\bigskip

The measurements proposed here are not all easy. The $B \to \pi K$ decays
should be characterized by branching ratios of order $10^{-5}$ for charged
pions and about half that for neutral pions if the $B \to \pi^- K^+$ decay
really has been observed at the $10^{-5}$ level \cite{Kpisep} and if the
gluonic penguin amplitude is dominant. The amplitudes in Fig.~\ref{piKquad}
are drawn to scale using the calculations of Ref.~\cite{DH}, neglecting
strong final-state phase differences, and assuming $\gamma = \pi/2$. The
effects of electroweak penguins can be seen not only in the rotation of the
phase of $A_{3/2}$ from its non-penguin value, but in substantial
differences in the lengths of the sides of the quadrangle. It may well be
that electroweak penguin effects make their first appearance in such rate
differences, as mentioned at the end of Sec.~III.

The $B_s \to \pi^0 \eta$ decay will be very difficult to measure. The
calculations of Ref.~\cite{DH} indicate a branching ratio of a couple of
parts in $10^7$. One has to distinguish a $B_s$ from a $\bar B_s$. In order
to observe the $\pi^0 \eta$ decay at a hadron machine, where the displaced
vertex of the $B_s$ would seem to be a prerequisite, one would have to
observe the $\eta$ in a mode involving charged particles.

Somewhat more hope is offered in the corresponding $B \to \pi K^*$ case, if
we can trust the very small branching ratio for $B_s \to \pi^0 \phi$ of a
couple of parts in $10^8$ predicted in Ref.~\cite{DHT}. (See also
\cite{Du}.) The corresponding electroweak penguin effects (characterizing
the diagonal (e) in Fig.~\ref{piKquad}) are expected to be smaller here,
whereas it is quite likely that the basic $B \to \pi K^*$ decays can be
observed soon.

The possibility of degeneracies in lengths of the sides of the quadrangles
can lead to a large amplification of errors in the amplitudes (e) when used
to predict the length of side (f). For example, imagine that (e) were
really zero and (a) = (c), (b) = (d). The length of (f) then would be
indeterminate.
On the other hand, if the diagonal (e) of the quadrangle is sufficiently
small, the quadrangle reduces to two nearly degenerate triangles in which
the effects of electroweak penguins are negligible.
In this case, the second diagonal is given to a good approximation by
$\sqrt{2}A(B^+\to\pi^0\rho^+)$ [assuming some cancellation between the
$P_P$ and $P_V$ terms of Eq.~(\ref{pirhoamp})], and the relative phase
between this amplitude and its charge-conjugate measures $2\gamma$.
Indeed, the very small value of ${\rm BR}(B_s\to\pi^0 \phi)$ calculated in
Ref.~\cite{DHT} suggests that this may be happening for the decays $B \to
\pi K^*$.

\bigskip
\centerline{\bf VI. CONCLUSIONS}
\bigskip

We have found the following results.

(a) Electroweak penguins (EWP's) are not expected to substantially affect
the discussion in Ref.~\cite{GL} regarding $B \to \pi \pi$ decays.

(b) EWP's {\em are} more likely to be important in the comparisons
\cite{BPP,PRL,PLB} of $B \to \pi K$ and $B \to \pi \pi$ decays, though such
conclusions are dependent on the evaluation of hadronic matrix elements of
operators.

(c) EWP's do not introduce new amplitudes of flavor SU(3), so that one
cannot detect their presence merely by modification of flavor-SU(3)
amplitude relations.

(d) A deviation of the rate ratio
$2 \Gamma(B^+\to\pi^0 K^+)/\Gamma(B^0\to \pi^- K^+)$ from unity indicates
the presence of EWP's, and similarly for
$2 \Gamma(B^0\to\pi^0 K^0)/\Gamma(B^+\to \pi^+ K^0)$. Since all of these
branching ratios are expected to be ${\cal O}(10^{-5})$, these are likely
to be the first (indirect) experimental signals of EWP's. Electroweak
penguins are expected to dominate decays of the form $B_s \to (\phi~{\rm
or}~\eta) + (\pi~{\rm or}~\rho)$ \cite{DHT}, but the branching ratios for
these processes are expected to be significantly smaller.

(e) A quadrangle analysis has been presented for such decays as $B \to \pi
K$, $B \to \pi K^*$, and $B \to \rho K$. One diagonal of the quadrangle is
related to the amplitude for a physical process such as $B_s \to \pi^0
\eta$ or $B_s \to \pi^0 \phi$, so that one can perform a construction to
obtain the other diagonal. From the magnitude and phase of this amplitude,
one can obtain $\sin \gamma$, where $\gamma \equiv {\rm Arg} (V^*_{ub})$.

(f) The $B \to \pi K^*$ processes hold out hope for a small electroweak
penguin contribution, if the $B_s \to \pi^0 \phi$ branching ratio is as
small as cited in Ref.~\cite{DHT}. In such a case, the quadrangle will
degenerate into two nearly identical triangles, so that the original
analysis of Ref.~\cite{PRL}, suitably modified to take account of the
presence of one vector and one pseudoscalar meson, may be more trustworthy.
We have presented the ingredients of such an analysis in Sec. IV C.

\bigskip
\centerline{\bf ACKNOWLEDGMENTS}
\bigskip

We thank J. Cline, A. Dighe, I. Dunietz, G. Eilam, A. Grant, K. Lingel, H.
Lipkin, R. Mendel, S. Stone, L. Wolfenstein, and M. Worah for fruitful
discussions. J. Rosner wishes to acknowledge the hospitality of the
Fermilab theory group and the Cornell Laboratory for Nuclear Studies during
parts of this investigation. M. Gronau, O. Hern\'andez and D. London are
grateful for the hospitality of the University of Chicago, where part of
this work was done. This work was supported in part by the United States --
Israel Binational Science Foundation under Research Grant Agreement
90-00483/3, by the German-Israeli Foundation for Scientific Research and
Development, by the Fund for Promotion of Research at the Technion, by the
NSERC of Canada and les Fonds FCAR du Qu\'ebec, and by the United States
Department of Energy under Contract No. DE FG02 90ER40560.

\bigskip

\def \ajp#1#2#3{Am.~J.~Phys.~{\bf#1}, #2 (#3)}
\def \apny#1#2#3{Ann.~Phys.~(N.Y.) {\bf#1}, #2 (#3)}
\def \app#1#2#3{Acta Phys.~Polonica {\bf#1}, #2 (#3)}
\def \arnps#1#2#3{Ann.~Rev.~Nucl.~Part.~Sci.~{\bf#1}, #2 (#3)}
\def \cmp#1#2#3{Commun.~Math.~Phys.~{\bf#1}, #2 (#3)}
\def \cmts#1#2#3{Comments on Nucl.~Part.~Phys.~{\bf#1}, #2 (#3)}
\def \cn{Collaboration}
\def \corn93{{\it Lepton and Photon Interactions:  XVI International Symposium,
Ithaca, NY August 1993}, AIP Conference Proceedings No.~302, ed.~by P. Drell
and D. Rubin (AIP, New York, 1994)}
\def \cp89{{\it CP Violation,} edited by C. Jarlskog (World Scientific,
Singapore, 1989)}
\def \dpff{{\it The Fermilab Meeting -- DPF 92} (7th Meeting of the American
Physical Society Division of Particles and Fields), 10--14 November 1992,
ed. by C. H. Albright \ite~(World Scientific, Singapore, 1993)}
\def \dpf94{DPF 94 Meeting, Albuquerque, NM, Aug.~2--6, 1994}
\def \efi{Enrico Fermi Institute Report No. EFI}
\def \el#1#2#3{Europhys.~Lett.~{\bf#1}, #2 (#3)}
\def \f79{{\it Proceedings of the 1979 International Symposium on Lepton and
Photon Interactions at High Energies,} Fermilab, August 23-29, 1979, ed.~by
T. B. W. Kirk and H. D. I. Abarbanel (Fermi National Accelerator Laboratory,
Batavia, IL, 1979}
\def \hb87{{\it Proceeding of the 1987 International Symposium on Lepton and
Photon Interactions at High Energies,} Hamburg, 1987, ed.~by W. Bartel
and R. R\"uckl (Nucl. Phys. B, Proc. Suppl., vol. 3) (North-Holland,
Amsterdam, 1988)}
\def \ib{{\it ibid.}~}
\def \ibj#1#2#3{~{\bf#1}, #2 (#3)}
\def \ichep72{{\it Proceedings of the XVI International Conference on High
Energy Physics}, Chicago and Batavia, Illinois, Sept. 6--13, 1972,
edited by J. D. Jackson, A. Roberts, and R. Donaldson (Fermilab, Batavia,
IL, 1972)}
\def \ijmpa#1#2#3{Int.~J.~Mod.~Phys.~A {\bf#1}, #2 (#3)}
\def \ite{{\it et al.}}
\def \jmp#1#2#3{J.~Math.~Phys.~{\bf#1}, #2 (#3)}
\def \jpg#1#2#3{J.~Phys.~G {\bf#1}, #2 (#3)}
\def \lkl87{{\it Selected Topics in Electroweak Interactions} (Proceedings of
the Second Lake Louise Institute on New Frontiers in Particle Physics, 15--21
February, 1987), edited by J. M. Cameron \ite~(World Scientific, Singapore,
1987)}
\def \ky85{{\it Proceedings of the International Symposium on Lepton and
Photon Interactions at High Energy,} Kyoto, Aug.~19-24, 1985, edited by M.
Konuma and K. Takahashi (Kyoto Univ., Kyoto, 1985)}
\def \mpla#1#2#3{Mod.~Phys.~Lett.~A {\bf#1}, #2 (#3)}
\def \nc#1#2#3{Nuovo Cim.~{\bf#1}, #2 (#3)}
\def \np#1#2#3{Nucl.~Phys.~{\bf#1}, #2 (#3)}
\def \pisma#1#2#3#4{Pis'ma Zh.~Eksp.~Teor.~Fiz.~{\bf#1}, #2 (#3) [JETP Lett.
{\bf#1}, #4 (#3)]}
\def \pl#1#2#3{Phys.~Lett.~{\bf#1}, #2 (#3)}
\def \plb#1#2#3{Phys.~Lett.~B {\bf#1}, #2 (#3)}
\def \pr#1#2#3{Phys.~Rev.~{\bf#1}, #2 (#3)}
\def \pra#1#2#3{Phys.~Rev.~A {\bf#1}, #2 (#3)}
\def \prd#1#2#3{Phys.~Rev.~D {\bf#1}, #2 (#3)}
\def \prl#1#2#3{Phys.~Rev.~Lett.~{\bf#1}, #2 (#3)}
\def \prp#1#2#3{Phys.~Rep.~{\bf#1}, #2 (#3)}
\def \ptp#1#2#3{Prog.~Theor.~Phys.~{\bf#1}, #2 (#3)}
\def \rmp#1#2#3{Rev.~Mod.~Phys.~{\bf#1}, #2 (#3)}
\def \rp#1{~~~~~\ldots\ldots{\rm rp~}{#1}~~~~~}
\def \si90{25th International Conference on High Energy Physics, Singapore,
Aug. 2-8, 1990}
\def \slc87{{\it Proceedings of the Salt Lake City Meeting} (Division of
Particles and Fields, American Physical Society, Salt Lake City, Utah, 1987),
ed.~by C. DeTar and J. S. Ball (World Scientific, Singapore, 1987)}
\def \slac89{{\it Proceedings of the XIVth International Symposium on
Lepton and Photon Interactions,} Stanford, California, 1989, edited by M.
Riordan (World Scientific, Singapore, 1990)}
\def \smass82{{\it Proceedings of the 1982 DPF Summer Study on Elementary
Particle Physics and Future Facilities}, Snowmass, Colorado, edited by R.
Donaldson, R. Gustafson, and F. Paige (World Scientific, Singapore, 1982)}
\def \smass90{{\it Research Directions for the Decade} (Proceedings of the
1990 Summer Study on High Energy Physics, June 25 -- July 13, Snowmass,
Colorado), edited by E. L. Berger (World Scientific, Singapore, 1992)}
\def \stone{{\it B Decays}, edited by S. Stone (World Scientific, Singapore,
1994)}
\def \tasi90{{\it Testing the Standard Model} (Proceedings of the 1990
Theoretical Advanced Study Institute in Elementary Particle Physics, Boulder,
Colorado, 3--27 June, 1990), edited by M. Cveti\v{c} and P. Langacker
(World Scientific, Singapore, 1991)}
\def \yaf#1#2#3#4{Yad.~Fiz.~{\bf#1}, #2 (#3) [Sov.~J.~Nucl.~Phys.~{\bf #1},
#4 (#3)]}
\def \zhetf#1#2#3#4#5#6{Zh.~Eksp.~Teor.~Fiz.~{\bf #1}, #2 (#3) [Sov.~Phys. -
JETP {\bf #4}, #5 (#6)]}
\def \zpc#1#2#3{Zeit.~Phys.~C {\bf#1}, #2 (#3)}

\end{document}